\documentclass[manuscript,screen,nonacm]{acmart}
\usepackage{subfigure}
\AtBeginDocument{%
  \providecommand\BibTeX{{%
    \normalfont B\kern-0.5em{\scshape i\kern-0.25em b}\kern-0.8em\TeX}}}

\setcopyright{acmcopyright}
\copyrightyear{2023}
\acmYear{2023}
\acmDOI{XXXXXXX.XXXXXXX}

\acmConference[Conference acronym 'XX]{Make sure to enter the correct
  conference title from your rights confirmation emai}{June 19,
  2023}{Online}
\acmBooktitle{The 1st International Workshop on Explainable AI for the Arts (XAIxArts), ACM Creativity and Cognition (C\&C).
2023, Online} 
\acmPrice{15.00}
\acmISBN{978-1-4503-XXXX-X/18/06}

\begin{document}

\title{An Autoethnographic Exploration of XAI in Algorithmic Composition}

\author{Ashley Noel-Hirst}
\email{a.l.noel-hirst@qmul.ac.uk}
\orcid{0009-0001-7987-9905}
\author{Nick Bryan-Kinns}
\email{n.bryan-kinns@qmul.ac.uk}
\orcid{0000-0002-1382-2914}
\affiliation{%
  \institution{Queen Mary University of London}
  \streetaddress{Mile End Rd, Bethnal Green}
  \city{London}
  \country{UK}
  \postcode{E1 4NS}
}


\fancyhead[LE, RO]{\footnotesize The 1st International Workshop on Explainable AI for the Arts (XAIxArts)}
\renewcommand{\shortauthors}{Noel-Hirst and Bryan-Kinns}

\begin{abstract}
Machine Learning models are capable of generating complex music across a range of genres from folk to classical music. However, current generative music AI models are typically difficult to understand and control in meaningful ways. Whilst research has started to explore how explainable AI (XAI) generative models might be created for music, no generative XAI models have been studied in music making practice. This paper introduces an autoethnographic study of the use of the MeasureVAE generative music XAI model with interpretable latent dimensions trained on Irish folk music.  Findings suggest that the exploratory nature of the music-making workflow foregrounds musical features of the training dataset rather than features of the generative model itself. The appropriation of an XAI model within an iterative workflow highlights the potential of XAI models to form part of a richer and more complex workflow than they were initially designed for. 





\end{abstract}

\ccsdesc[500]{Human-centered computing~Interactive systems and tools}
\ccsdesc[300]{Human-centered computing~Graphical user interfaces}
\ccsdesc[500]{Applied computing~Sound and music computing}

\keywords{ethnographic artificial intelligence, music generation, explainable artificial intelligence}



\maketitle

\begin{small}

\textbf{ACM Reference Format: \\}
Ashley Noel-Hirst and Nick Bryan-Kinns. 2023. An Autoethnographic Exploration of XAI in Algorithmic Composition. In \textit{The 1st International Workshop on Explainable AI for the Arts (XAIxArts)}, ACM Creativity and Cognition (C\&C) 2023. Online, 3 pages. \url{https://xaixarts.github.io}
\\

\end{small}


\section{Introduction}

In recent years, computational approaches to music generation and creativity support have prompted and adopted a range of Artificial Intelligence (AI) techniques, especially Machine Learning. Composers who utilize novel musical representations, 
timbres, 
and pitch and time divisions 
go so far as to create new software for their works to be realized often using AI tools support this process \cite{lewis_interacting_1999}. These approaches broadly fall into two paradigms: the appropriation of materials from existing deep learning models \cite{knotts_ai-lectronica_2021}, and the development of novel models e.g. \cite{fiebrink_machine_2018}. 
However, regardless of the paradigm or application of AI for music, the underlying processes are typically hard to understand or control in meaningful ways \cite{bryan-kinns_exploring_nodate}. Whilst explainable AI (XAI) \cite{vilone_notions_2021} and Human-Centred AI (HCAI) \cite{shneiderman_human-centered_2020} aim to make AI models more understandable to humans, there have been no studies of how XAI approaches might be used in artistic practice.





\section{Exploring an Explainable Variational Autoencoder}

To explore the use of XAI in creative practice we report on an autoethnographic exploration of a generative music XAI system – MeasureVAE \cite{pati_latent_2019} implemented as a Max4Live plugin \cite{banar_tool_2023} - as an illustrative case study of how such XAI models might be appropriated in music-making practice. The autoethographic exploration is reported in first person by the first author whose music making practice involves composing interactive and generative musical systems 
\cite{clarice_hilton_figural_2023}
, which are then rendered as fixed electronic works. 

\begin{figure}[h]
  \centering
  \includegraphics[width=0.5\linewidth]{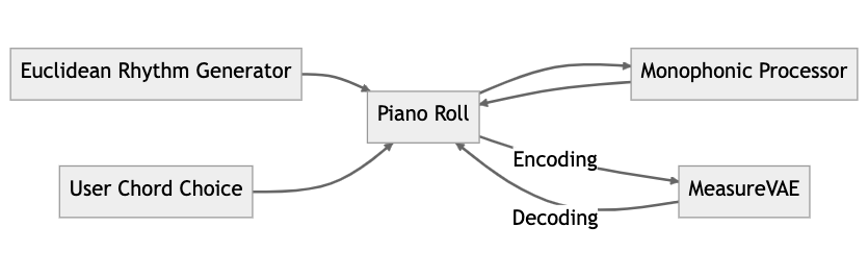}
  \caption{Musical workflow for generating musical measures from Euclidean Rhythms.}\label{fig:workflow}
  \Description{A map, of algorithmic procedures. Links include: Euclidean rhythm generator to piano roll; User chord choice to piano roll; Monophonic processor to and from the piano roll; MeasureVAE to and from the piano roll.}
\end{figure}

In my practice, I utilize a range of rule-based composition systems with other AI methods. These are typically focused on the evolution of rhythms over time, and integrate a number of Max/MSP tools – such as Euclidean sequencers, which generate complex rhythms from a few parameters. These rhythms are found in a number of folk traditions, but interestingly they are not represented in the Irish folk data that MeasureVAE was trained on. To explore this dissonance between musical practice and AI training data I reflexively iterated through several approaches to music making. Here I report on one approach informed by my music making practice in which Euclidean rhythms are used to drive and explore the generation of melodies with MeasureVAE – as illustrated by the workflow in figrue \ref{fig:workflow}.

Euclidean rhythms are lists of pulses [x] and rests [.]. These lists, expressed as E(i,j,k), describe i pulses dividing the j total beats as evenly as possible, displaced to the right by k positions. For example, the phrase [x..x.x..x.x..] is the Euclidean Rhythm expressed by E(5,13,0). 
Toussaint \cite{toussaint_euclidean_nodate} outlined how this algorithm can be used to generate, analyse and permute a range of rhythmic patterns from across the world. It can also be found in use in minimalist composition \cite{lawton_hall_euclidean_2020} as well as being used by live coders and modular synth users \cite{roberts_blawans_2021}.

\begin{figure}[h]
  \centering
  \begin{subfigure}[]
        \centering
        \includegraphics[width=0.23\linewidth]{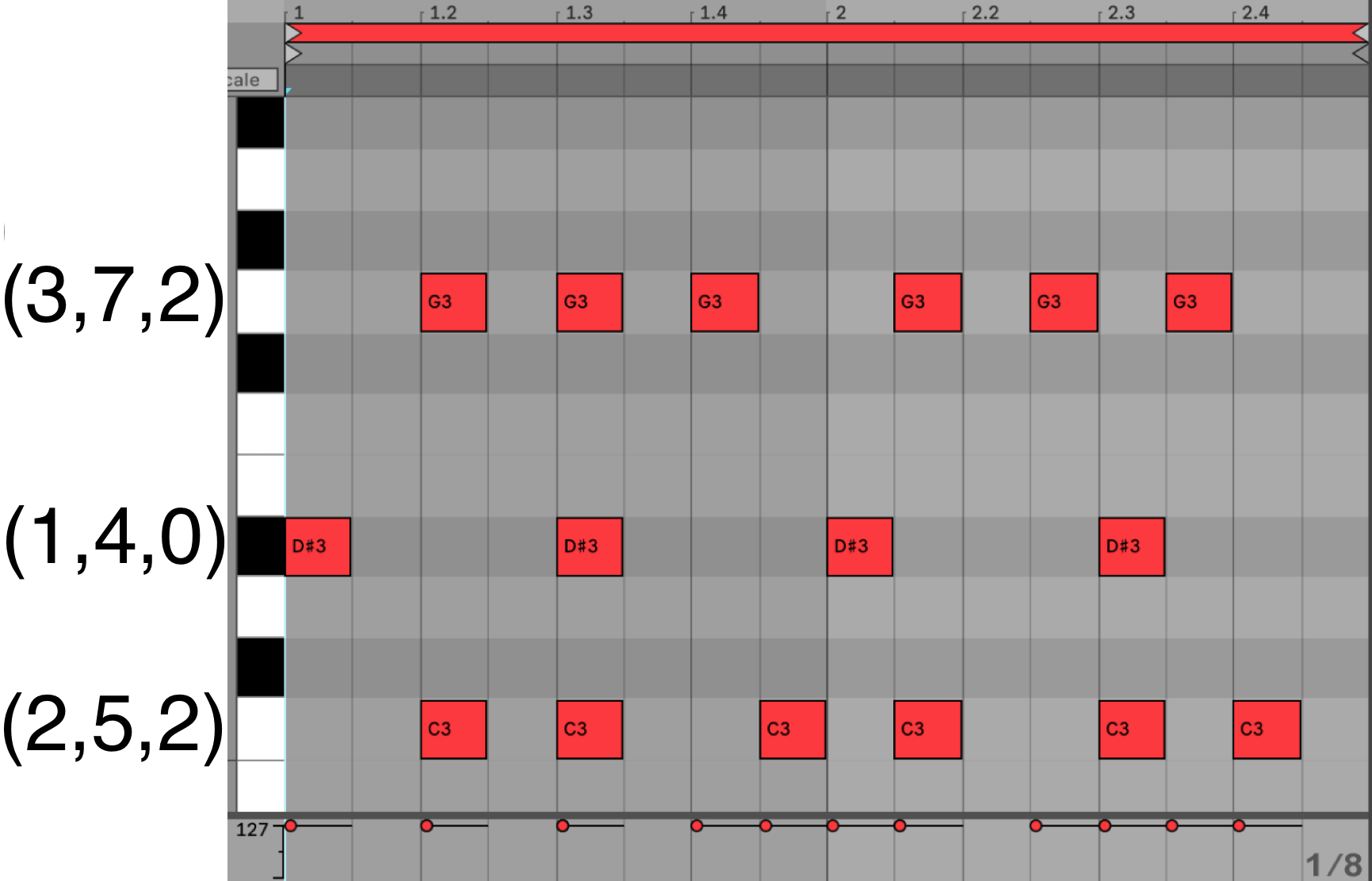}
        \end{subfigure}
  \hfill
    \begin{subfigure}[]
        \centering
        \includegraphics[width=0.212\linewidth]{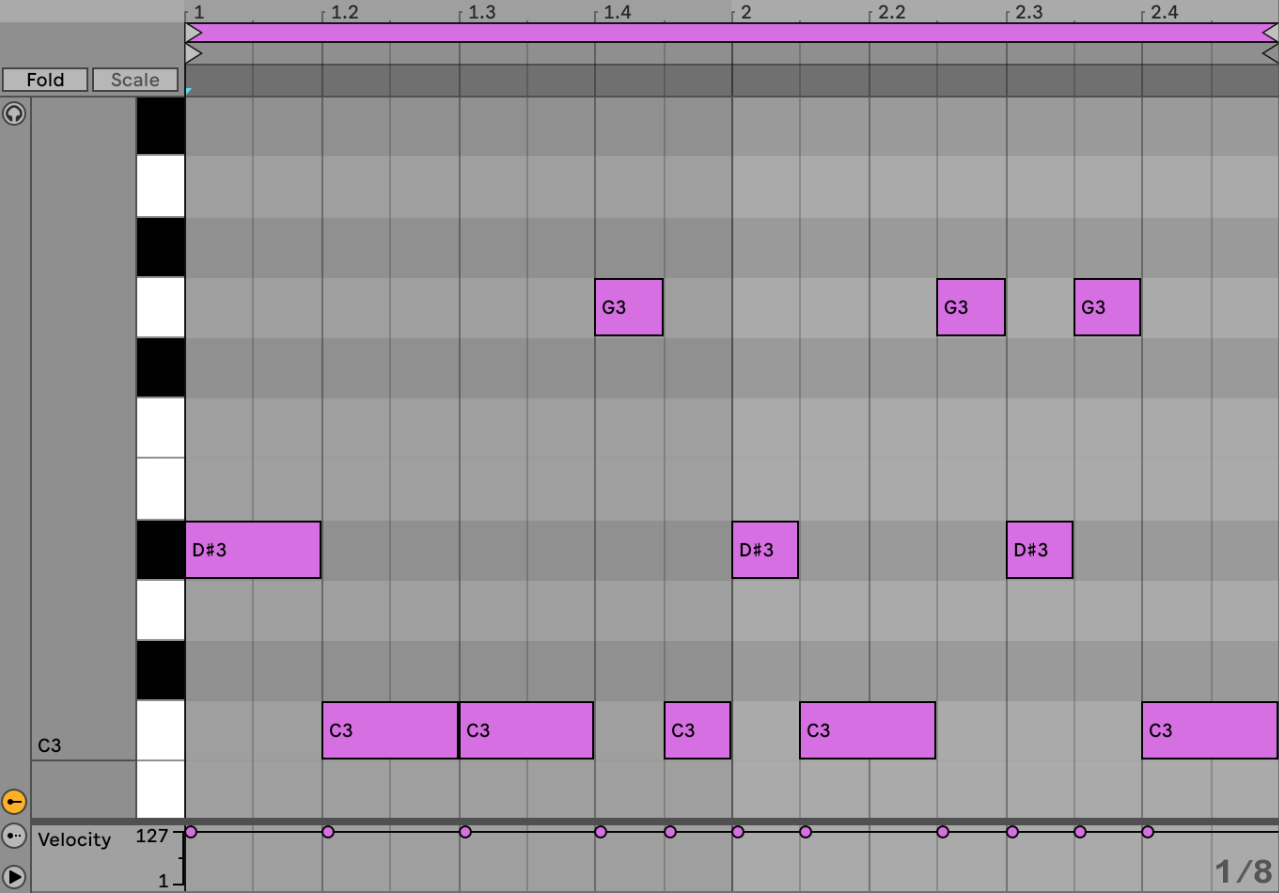}
        \end{subfigure}
    \hfill
    \begin{subfigure}[]
        \centering
        \includegraphics[width=0.2\linewidth]{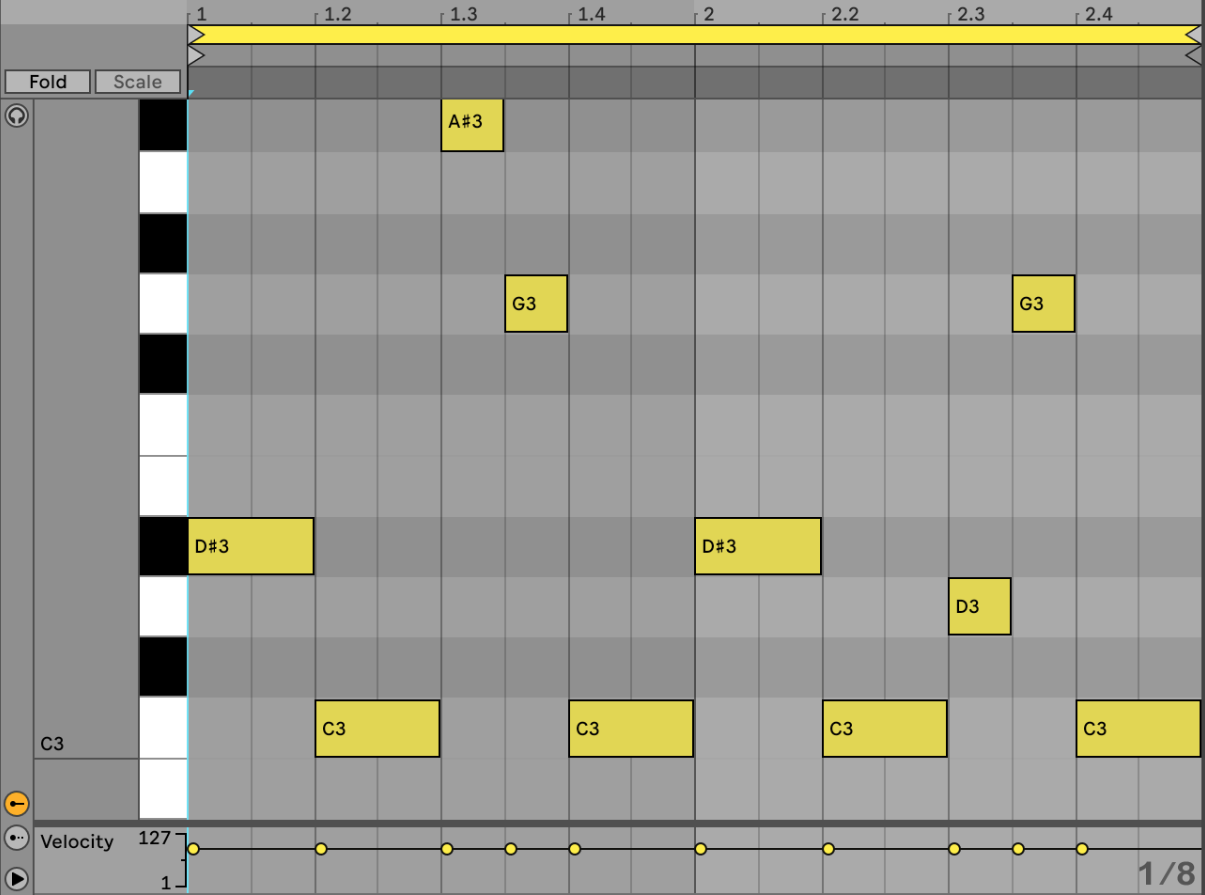}
        \end{subfigure}
        \hfill
    \begin{subfigure}[]
        \centering
        \includegraphics[width=0.2\linewidth]{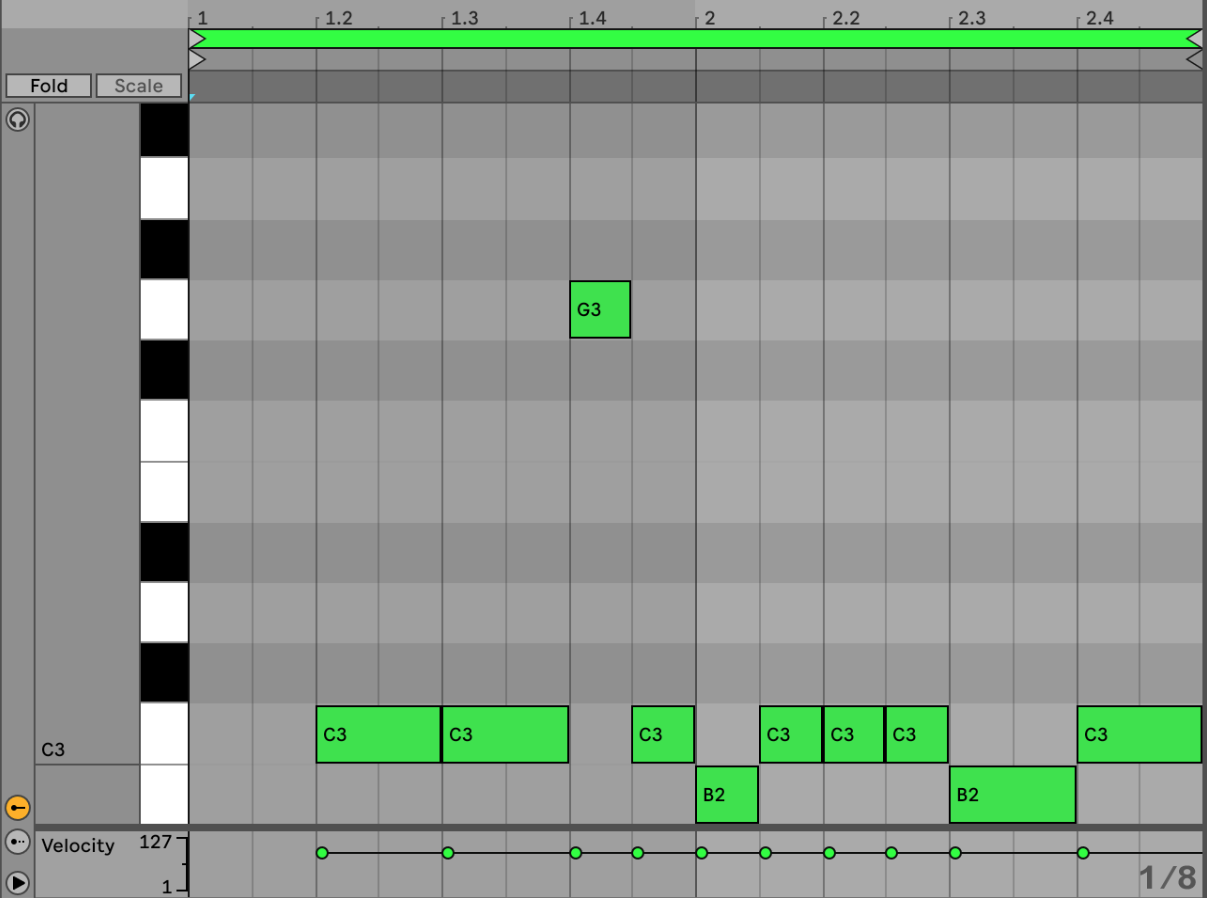}
        \Description{}
        \end{subfigure}
        \caption{a) Three Euclidean rhythms are played in parallel using notes from a C minor chord; b) This is made monophonic, generating a melody; c) Its encoding without any modification; d) Its encoding after the Euclidean parameters were altered.}
        \label{fig:melodies}
\end{figure}

In my own practice, I typically use a range of parameterised sequences in parallel to drive the creation of new musical phrases. For this study I created a number of Euclidean sequences: E(3,7,2) [. . X . X . X]; E(4,16,0) [X . . .]; and E(2,5,2) [. . X . X]. I then applied each rhythm to a note from a C minor chord, resulting in the poly-rhythm-based melody illustrated in figure \ref{fig:melodies}a.
Since MeasureVAE needs monophonic input, monophonic melodies were extracted using a lowest note policy – restraint brings tension to the infrequent high notes. When applied to the pattern in figure \ref{fig:melodies}a, this produces the legato melody in figure \ref{fig:melodies}b. This monophonic melody is then fed in to MeasureVAE which generates an output illustrated in figure \ref{fig:melodies}c. Permuting the parameters of each Euclidean rhythm produces a number of similar melodies with comparable divergence from the model. I then explored modulations of my Euclidean system, composing `into' the fixed MeasureVAE to produce new melodies as illustrated in figure \ref{fig:melodies}d. Through this process I found musical artifacts in the MeasureVAE – encoded elements which reflected the underlying training data (or lack thereof) more than the input itself.




\subsection{Feedback and Explainability}

Through exploration, I found that employing MeasureVAE in tandem with other computational composition systems extended MeasureVAE's explainability beyond its intended scope. First, we gain new avenues for explainability. MeasureVAE provides a navigable heatmap of the latent space, highlighting areas of the source distribution which are more or less dense in the original dataset. In the denser areas, regularisation makes for useful `contrastive feature' explanation and navigation.
By feeding a series of permuting Euclidean sequences into MeasureVAE, 
I was able to explore the under-defined contours of the latent space through example. My parametrically-defined melodies would be approximated within the distribution of the original dataset, and the artifacts which arose during this process gave some idea of the extent to which my melody was congruent with the distribution of the training set. Conversely, this could also be thought of as an evaluation of the appropriateness of the original dataset for my task, or a qualitative understanding of the high dimensional topology of the latent space, going beyond the 4 regularised dimensions originally explored \cite{bryan-kinns_exploring_nodate}. 

Second, there is some appropriation of the explainability of one system to another. For example, Euclidean systems are difficult to navigate and evaluative feedback from MeasureVAE in the music making process helped to inform the tweaking of my Euclidean system. Specifically, the MeasureVAE encoder gave me feedback about the extent to which my melody caused activation in its regularised dimensions. With this, I was able to more fluidly modify my Euclidean parameters so as to create melodies which were encoded as being less dense or complex e.g. \ref{fig:melodies}d. 

\section{Conclusions}

Explainable AI music systems promise increased control and transparency to the computer music composer. As illustrated in this paper, the location of the explainability in a musical workflow has implications for the types of feedback it can give - in our case, exploring the features of the training set, and appropriating explainability from an XAI to an algorithmic composition process. Future work includes exploring how we can leverage such re-contextualised feedback for algorithmic surprise, qualitative evaluation of datasets, and cross-genre adaptation of pre-trained models.






\bibliographystyle{ACM-Reference-Format}
\bibliography{sample-base}

\end{document}